\documentclass{Odyssey2026}

\interspeechcameraready 

\usepackage{multirow}
\usepackage{booktabs}
\usepackage{float}

\usepackage{numprint}
\npdecimalsign{.}
\newcommand{\eer}[1]{\nprounddigits{2}\numprint{#1}}
\newcommand{\dcf}[1]{\nprounddigits{3}\numprint{#1}}

\newcommand{\paragraphTwo}[1]{%
  \vspace{6pt}
  \noindent
  \textbf{#1}
}

\title{Assessing the Energy and Carbon Emissions of Neural Speaker Verification Model in Training and Inference}

\author[affiliation={1,2}]{Hugo}{Leguillier}
\author[affiliation={1}]{Driss}{Matrouf}
\author[affiliation={2}]{Guillaume}{Lechien}
\author[affiliation={1}]{Mickael}{Rouvier}

\affiliation{Avignon University}{LIA, UPR 4128}{France}
\email{name.surname@univ-avignon.fr}
\affiliation{}{Aday}{France}
\email{firstname.lastname@univ-avignon.fr, glechien@aday.fr}

\usepackage{comment}
\usepackage{soul}

\usepackage{booktabs}
\usepackage{multirow}
\usepackage{makecell}
\usepackage[table]{xcolor}
\usepackage{colortbl}

\begin{document}

\maketitle

\begin{abstract}
Deep-learning speaker verification (SV) increasingly relies on deep neural network backbones, whose environmental impact remains largely undocumented. In this paper, we conduct an evaluation of ResNet architectures trained on VoxCeleb2, varying depth, channel width, and stage distribution, and measure energy consumption and carbon footprint using node-level sensors. Results show a clear point of diminishing returns: deeper or wider models bring only marginal accuracy gains while energy consumption grows steeply. In contrast, mid-sized networks such as ResNet-50 and stage-concentrated variants achieve favorable trade-offs between performance and environmental impact. These findings provide actionable guidelines for designing energy-efficient SV systems.
\end{abstract}

\section{Introduction}
\label{sec:intro}

Speaker verification (SV) aims to verify a speaker's identity based solely on their voice, with applications ranging from forensic analysis~\cite{forensicSV} to multimedia ~\cite{gresse2017acoustic}.

Modern speaker verification systems rely on Deep Neural Networks (DNNs) to extract fixed-dimensional speaker embeddings. These embeddings, extracted from hidden layers, generalize well to unseen speakers and have achieved strong performance in major evaluations (NIST SRE~\cite{alamabc}, VoxCeleb~\cite{makarov2022id}). Over time, neural architectures such as TDNN, ResNet, ECAPA2, and ReDimNet have substantially increased in depth (from 9 to 293 layers), leading to significant performance gains. The intuition is that deeper networks can capture increasingly discriminative information, producing embeddings that are both robust and highly separable.

However, these performance gains do not come without trade-offs. As models grow deeper and larger, the increase in the number of parameters and arithmetic operations leads to substantially higher memory requirements, longer runtimes, and greater energy consumption, both during training and inference. This trend raises several challenges, including hardware constraints for embedded deployment, financial costs related to computing infrastructures, and the carbon footprint associated with electricity use~\cite{patterson2021carbon}. These concerns are further amplified by the extensive GPU resources required to train deep networks and by repeated training runs for hyperparameter tuning or transfer learning. Therefore, it is essential to quantify the trade-offs among model complexity, performance, energy consumption, and carbon footprint in order to guide the design of more sustainable speaker verification systems.

In this context, evaluating speaker verification systems solely in terms of performance metrics such as EER and minDCF is no longer sufficient. It is also necessary to account for their energy consumption and carbon footprint in order to better understand the trade-offs between accuracy, model complexity, and environmental impact.

In this paper, we systematically evaluate ResNet-based SV models by varying three key structural dimensions: \emph{depth} (from compact to very deep), \emph{width} (scaling channel counts), and \emph{stage distribution} (redistributing residual blocks across stages). Models are trained on VoxCeleb2 and tested both in-domain (VoxCeleb1-O, E, and H) and out-of-domain (CommonBench, CN-Celeb)~\cite{rusti2023benchmarkdatasetdynamicsbias}.

The main contributions of this paper are as follows:

\begin{itemize}

\item \textbf{Comparison of ResNet variants:} We conduct a comparison of ResNet variants by varying three key structural dimensions—depth, width, and residual block distribution—in order to precisely characterize the trade-offs between speaker verification performance and environmental impact.

\item \textbf{Evaluation across in-domain and out-of-domain datasets:} We evaluate these models on both in-domain and out-of-domain test sets using standard SV metrics (EER, minDCF), and relate these performance results to direct measurements of energy consumption and carbon footprint.

\item \textbf{Recommendations for energy-efficient SV design:} Based on these analyses, we provide concrete recommendations for the design of energy-efficient SV systems by identifying the configurations that offer the best trade-off between performance, energy consumption, and carbon footprint.

\end{itemize}

The paper is organized as follows: Section~\ref{sec:related_work} reviews the related work. Section~\ref{sec:quantifying_emissions} details the methodology for quantifying energy consumption and carbon emissions, specifying the chosen indicators (e.g., kWh, kg CO\textsubscript{2}eq) and the measurement tools employed. Section~\ref{sec:architecture_training_setup} introduces the ResNet architectures. Section~\ref{sec:experimental_setups} presents the experimental setup and results, including in-domain and out-of-domain performance, training energy consumption, and an analysis of the trade-offs between architectural choices, verification accuracy, and environmental impact. Finally, Section~\ref{sec:conclusions} concludes the paper.

\section{Related work}
\label{sec:related_work}

Recent work on sustainable machine learning has emphasized that model evaluation should not rely on accuracy alone, but should also account for energy consumption and carbon footprint. Early studies highlighted the financial cost and environmental impact of training large neural models and advocated for more transparent reporting of energy-related metrics \cite{strubell2019energy,anthony2020carbontracker}. Subsequent work further showed that the carbon footprint of deep learning depends not only on model size, but also on hardware efficiency, datacenter infrastructure, geographic location, and time-varying carbon intensity of electricity \cite{patterson2021carbon,dodge2022measuring}.

In speech processing, explicit analyses of energy consumption and carbon footprint remain comparatively limited. Parcollet and Ravanelli \cite{parcollet2021energy} quantified the energy consumption and carbon footprint of training end-to-end ASR systems, showing that small performance gains may come at a disproportionately high environmental impact. Dinarelli et al. \cite{dinarelli2022lowcost} studied a similar trade-off in end-to-end spoken language understanding, relating model performance to training time and electricity consumption. More recently, LeBenchmark 2.0 \cite{parcollet2024lebenchmark2} highlighted the environmental impact of large-scale self-supervised speech pre-training, showing that increasing the amount of pre-training data can substantially raise energy consumption and carbon footprint. Kulkarni et al. \cite{kulkarni2024sustainability} extended this perspective to large ASR systems at inference time by analyzing the energy consumption and carbon footprint of Whisper and MMS across multiple GPU platforms.

However, to the best of our knowledge, there is currently no published study specifically dedicated to analyzing the trade-off between speaker verification performance, energy consumption, and carbon footprint. Our work addresses this gap by focusing on ResNet-based SV systems and by studying how depth, width, and stage distribution affect the balance between accuracy and environmental impact.

\section{Environmental Impact Indicators and Measurement Methodology}
\label{sec:quantifying_emissions}

\subsection{Environmental indicators}
\label{sec:selected_indicators}

In this study, two indicators of environmental impact are considered: energy consumption and carbon footprint.

\paragraphTwo{Energy consumption}: measured in kilowatt-hours (kWh), this indicator reflects the total electricity consumption of a process. In practice, it mainly comes from GPU and CPU usage~\cite{nana2023energyconcernshpcsystems}. While hardware power ratings provide rough estimates, actual consumption depends on workload, utilization, memory access, and software efficiency~\cite{delestrac:hal-04604802}. Direct kWh measurement therefore offers a more accurate assessment during training.

\paragraphTwo{Carbon footprint}: expressed in kilograms (or grams) of CO\textsubscript{2} equivalent (CO\textsubscript{2}e), this metric represents the total greenhouse-gas impact associated with the electricity consumed in computational tasks. It depends on the carbon intensity of local energy sources, each of which has a specific emission factor derived from life-cycle analyses~\cite{dodge2022measuringcarbonintensityai}. In practice, the carbon footprint is estimated by converting energy consumption (kWh) into CO\textsubscript{2}eq using regional conversion factors published by environmental agencies.

These two indicators provide a coherent framework for assessing the environmental impact of model training and inference.

\subsection{Measurement tools}
\label{sec:measurement_tools}

Energy consumption and carbon footprint are tracked using the Compute Energy and Emissions Monitoring Stack (CEEMS)~\cite{Paipuri_2024}, an open-source, platform-independent tool specifically developed to monitor energy usage and emissions of individual workloads on high-performance computing and cloud infrastructures.

CEEMS gathers fine-grained energy consumption data directly from node-level hardware sensors, specifically through two primary measurement mechanisms:

\begin{itemize}
\item \textbf{Intel's Running Average Power Limit (RAPL)}: this mechanism reports granular power metrics specifically for CPU and DRAM components at microsecond intervals, providing insights into computational resource consumption.
\item \textbf{Baseboard Management Controllers via the IPMI-DCMI}: this interface provides node-level power measurements at lower frequency, covering additional system components beyond the CPU and DRAM.
\end{itemize}

Regarding carbon footprint estimation, CEEMS converts measured electricity consumption (in kWh) into CO\textsubscript{2}eq using real-time and static emission factors. Real-time emission data are retrieved from trusted sources like RTE eCO\textsubscript{2} \cite{rte2025eco2mix} (the French electricity transmission system operator) and Electricity Maps \cite{electricitymaps2025}, providing accurate and dynamic grid-emission intensities. Additionally, static emission factors from authoritative sources such as Our World in Data (OWID) are used as reference benchmarks.

\section{Speaker Embedding Extraction with ResNet}
\label{sec:architecture_training_setup}

Many speaker verification (SV) systems are currently based on the ResNet architecture or its variants (e.g., Res2Net, ResNeXt, and ECAPA-style backbones). ResNet is particularly well suited to SV because it combines strong representational capacity with a flexible design that can be adapted to different computational budgets. Its architecture is organized into four stages, each composed of several residual blocks. Within each block, convolutional layers are followed by batch normalization and nonlinear activation functions, while a shortcut connection adds the input directly to the transformed output. This residual learning mechanism facilitates the optimization of deep networks by preserving gradient flow during backpropagation and enabling the training of substantially deeper models.

A major advantage of ResNet lies in its configurability. The architecture can be scaled in depth by varying the number of residual blocks, as illustrated in configurations ranging from 18 to 419 layers, and in width by adjusting the number of feature maps through width multipliers. In speaker verification, the convolutional backbone first produces frame-level representations from acoustic features such as filterbanks. These representations are then aggregated over time by a pooling layer to form a fixed-dimensional utterance-level representation. This aggregation step is essential because speech utterances are variable in duration, whereas the downstream scoring pipeline requires a compact and fixed-size embedding. The architecture is typically completed with a fully connected layer that maps the pooled representation into a discriminative speaker embedding space.

ResNet is trained on a speaker identification task, where the objective is to classify each speech segment into one of $n$ known speaker classes in the training set. Once training is completed, the final classification layer is discarded, and a hidden layer at the utterance level is used to extract the speaker representation, commonly referred to as the speaker embedding. These embeddings are designed to capture speaker-specific characteristics while being robust to nuisance factors such as channel variability, background noise, or phonetic content. In practice, speaker embeddings can then be compared using similarity measures or back-end classifiers in order to perform verification.

\begin{table*}[t]
\centering
\setlength{\tabcolsep}{4pt}
\renewcommand{\arraystretch}{1.12}
\resizebox{\textwidth}{!}{%
\begin{tabular}{@{}lccccc@{\hspace{8pt}}cc cc cc @{\hspace{8pt}}cc cc@{}}
\toprule

\multirow{3}{*}{\textbf{System}} 
& \multirow{3}{*}{\makecell{\textbf{Params}\\\textbf{(M)}}}
& \multirow{3}{*}{\makecell{\textbf{Avg.}\\\textbf{EER}}}
& \multirow{3}{*}{\makecell{\textbf{Avg.}\\\textbf{minDCF}}}
& \multicolumn{2}{c}{\textbf{Environmental}} 
& \multicolumn{6}{c}{\textbf{In-domain}} 
& \multicolumn{4}{c}{\textbf{Out-of-domain}} \\

\cmidrule(lr){5-6} \cmidrule(lr){7-12} \cmidrule(l){13-16}

& & & 
& \multirow{2}{*}{\makecell{\textbf{Energy}\\\textbf{(kWh)}}}
& \multirow{2}{*}{\makecell{\textbf{CO$_2$}\\\textbf{(kg-eq)}}}
& \multicolumn{2}{c}{\textbf{Vox1-O}}
& \multicolumn{2}{c}{\textbf{Vox1-E}}
& \multicolumn{2}{c}{\textbf{Vox1-H}}
& \multicolumn{2}{c}{\textbf{CommonBench}}
& \multicolumn{2}{c}{\textbf{CN-Celeb}} \\

\cmidrule(lr){7-8} \cmidrule(lr){9-10} \cmidrule(lr){11-12} \cmidrule(lr){13-14} \cmidrule(l){15-16}

& & & & & 
& \textbf{EER} & \textbf{minDCF}
& \textbf{EER} & \textbf{minDCF}
& \textbf{EER} & \textbf{minDCF}
& \textbf{EER} & \textbf{minDCF}
& \textbf{EER} & \textbf{minDCF} \\

\midrule

\multicolumn{16}{@{}l}{\textit{Depth scaling}} \\
\midrule

ResNet-419-D & 133 & \textbf{\eer{3.349}} & \textbf{\dcf{0.218}} & 895.67 & 13.734 
& \textbf{\eer{0.680}} & \textbf{\dcf{0.052}} 
& \textbf{\eer{0.799}} & \textbf{\dcf{0.083}} 
& \eer{1.420} & \textbf{\dcf{0.135}} 
& \eer{3.475} & \dcf{0.330} 
& \textbf{\eer{10.374}} & \textbf{\dcf{0.492}} \\

ResNet-200-D & 61 & \eer{3.443} & \dcf{0.224} & 222.53 & 3.091 
& \eer{0.715} & \dcf{0.053} 
& \eer{0.819} & \dcf{0.083} 
& \textbf{\eer{1.418}} & \dcf{0.136} 
& \textbf{\eer{3.456}} & \dcf{0.329} 
& \eer{10.807} & \dcf{0.520} \\

ResNet-101-D & 38 & \eer{3.543} & \dcf{0.229} & 135.27 & 4.263 
& \eer{0.710} & \dcf{0.067} 
& \eer{0.820} & \dcf{0.087} 
& \eer{1.447} & \dcf{0.136} 
& \eer{3.502} & \textbf{\dcf{0.326}} 
& \eer{11.236} & \dcf{0.528} \\

ResNet-50-D & 17 & \eer{3.7508} & \dcf{0.245} & 51.69 & 0.982 
& \eer{0.877} & \dcf{0.078} 
& \eer{0.921} & \dcf{0.099} 
& \eer{1.624} & \dcf{0.157} 
& \eer{3.814} & \dcf{0.345} 
& \eer{11.518} & \dcf{0.548} \\

ResNet-34-D & 16 & \eer{3.899} & \dcf{0.254} & 46.2 & 0.661 
& \eer{0.989} & \dcf{0.082} 
& \eer{0.997} & \dcf{0.103} 
& \eer{1.747} & \dcf{0.169} 
& \eer{3.935} & \dcf{0.359} 
& \eer{11.827} & \dcf{0.555} \\

ResNet-18-D & 9 & \eer{4.108} & \dcf{0.273} & 63.05 & 0.790 
& \eer{1.191} & \dcf{0.103} 
& \eer{1.193} & \dcf{0.124} 
& \eer{2.041} & \dcf{0.194} 
& \eer{4.181} & \dcf{0.373} 
& \eer{11.940} & \dcf{0.571} \\

\midrule

\multicolumn{16}{@{}l}{\textit{Stage-distribution scaling}} \\
\midrule

ResNet-50-D$_{[8,2,2,2]}$ & 11 & \eer{3.801} & \dcf{0.248} & 76.67 & 1.127 
& \eer{0.965} & \dcf{0.0858} 
& \eer{0.978} & \dcf{0.1011} 
& \eer{1.669} & \dcf{0.1604} 
& \eer{3.809} & \dcf{0.346} 
& \eer{11.585} & \dcf{0.5476} \\

ResNet-50-D$_{[2,8,2,2]}$ & 11 & \eer{3.651} & \dcf{0.244} & 52.06 & 0.769 
& \eer{0.834} & \dcf{0.088} 
& \eer{0.928} & \dcf{0.0958} 
& \eer{1.596} & \dcf{0.1554} 
& \eer{3.742} & \dcf{0.347} 
& \eer{11.156} & \dcf{0.5375} \\

ResNet-50-D$_{[2,2,8,2]}$ & 17 & \eer{3.684} & \dcf{0.242} & 49.33 & 0.728 
& \eer{0.842} & \dcf{0.0718} 
& \eer{0.903} & \dcf{0.0943} 
& \eer{1.571} & \dcf{0.153} 
& \eer{3.710} & \dcf{0.347} 
& \eer{11.393} & \dcf{0.544} \\

ResNet-50-D$_{[2,2,2,8]}$ & 17 & \eer{3.906} & \dcf{0.260} & 47.0 & 0.688 
& \eer{0.930} & \dcf{0.0853} 
& \eer{1.032} & \dcf{0.1111} 
& \eer{1.813} & \dcf{0.172} 
& \eer{4.020} & \dcf{0.37} 
& \eer{11.737} & \dcf{0.563} \\

ResNet-50-D$_{[2,6,6,2]}$ & 16 & \eer{3.645} & \dcf{0.233} & 53.82 & 0.796 
& \eer{0.760} & \dcf{0.054} 
& \eer{0.878} & \dcf{0.0917} 
& \eer{1.523} & \dcf{0.146} 
& \eer{3.657} & \dcf{0.343} 
& \eer{11.410} & \dcf{0.533} \\

\midrule

\multicolumn{16}{@{}l}{\textit{Width scaling}} \\
\midrule

ResNet-50-W4 & 233 & \eer{3.767} & \dcf{0.246} & 438.46 & 6.98 
& \eer{0.821} & \dcf{0.0876} 
& \eer{0.874} & \dcf{0.089} 
& \eer{1.593} & \dcf{0.1485} 
& \eer{3.919} & \dcf{0.358} 
& \eer{11.63} & \dcf{0.545} \\

ResNet-50-W2 & 61 & \eer{3.680} & \dcf{0.23574} & 120.76 & 1.66 
& \eer{0.826} & \dcf{0.0619} 
& \eer{0.849} & \dcf{0.089} 
& \eer{1.536} & \dcf{0.1468} 
& \eer{3.839} & \dcf{0.353} 
& \eer{11.34} & \dcf{0.528} \\

ResNet-50-W0.5 & 5 & \eer{3.885} & \dcf{0.257} & 40.05 & 0.62 
& \eer{1.005} & \dcf{0.0837} 
& \eer{1.057} & \dcf{0.108} 
& \eer{1.800} & \dcf{0.1738} 
& \eer{3.867} & \dcf{0.349} 
& \eer{11.698} & \dcf{0.571} \\

ResNet-50-W0.25 & 2 & \eer{4.581} & \dcf{0.299} & \textbf{34.35} & \textbf{0.47} 
& \eer{1.488} & \dcf{0.1431} 
& \eer{1.528} & \dcf{0.155} 
& \eer{2.501} & \dcf{0.2263} 
& \eer{4.445} & \dcf{0.386} 
& \eer{12.942} & \dcf{0.587} \\

\bottomrule
\end{tabular}%
}
\caption{Speaker verification performance (EER \% / minDCF) and environmental impact indicators (energy consumption in kWh and carbon footprint in kg CO$_2$eq) of different ResNet architectures across in-domain and out-of-domain benchmarks.}
\label{tab:Performance_EER}
\end{table*}

\section{Experiments and results analysis}
\label{sec:experimental_setups}

\subsection{Experimental setup}
\label{sec:experimental_setup_detail}

For these experiments, we use the Kiwano toolkit~\cite{rouvier2026}\footnote{\url{https://github.com/kiwano-toolkit/kiwano}} and train ResNet-based speaker embedding extractors on VoxCeleb2~\cite{chung2018voxceleb2}. Training uses mini-batches of 512 and 3.5\,s random crops. Standard data augmentation follows~\cite{snyder2018x} with MUSAN~\cite{snyder2015musanmusicspeechnoise}, RIRs~\cite{rirs} and SpecAugment~\cite{park2019specaugment}. Inputs are 80-dimensional filterbanks, and embeddings are 256-dimensional. Models are trained for 42 epochs with AAM loss, SGD (momentum 0.9, weight decay $2\!\times\!10^{-4}$). Squeeze-and-Excitation is enabled on the first two stages~\cite{rouvier2021studyingsqueezeandexcitationusedcnn}; the four stages use \{128,128,256,256\} feature maps.

We trained different ResNet models by adjusting depth, width, and stage distribution.

\paragraphTwo{Depth Scaling}: These models, referred to as \emph{ResNet-X-D}, increase the number of layers (depth) in the network. The depth is denoted by $X$, which can take values such as 18, 34, 50, 101, 200, or 419. All variants share the same ResBlock design; their only difference lies in how many blocks are assigned to each stage, which directly determines the overall depth.

\paragraphTwo{Width Scaling}: These models, referred to as \emph{ResNet-50-W$x$}, adjust the number of feature maps in each layer by a constant factor $x$. For instance, the baseline \emph{ResNet-50-D} uses feature maps $\{128,128,256,256\}$ across its four stages. In contrast, \emph{ResNet-50-W2} doubles these values to $\{256,256,512,512\}$, while \emph{ResNet-50-W0.5} reduces them to $\{64,64,128,128\}$.

\paragraphTwo{Stage Distribution Scaling}: These models, referred to as \emph{ResNet-50-D$_{[a,b,c,d]}$}, keep the total depth fixed (e.g., 50 layers) but redistribute residual blocks across stages. For example, in an early-concentrated variant \emph{ResNet-50-D$_{[8,2,2,2]}$}, most blocks are shifted to the first stage, while later stages contain fewer residual blocks. Because the stages do not have the same width, this redistribution also changes the parameter count and the computational profile of the network.


We evaluate in-domain on VoxCeleb1-O/E/H (cleaned)~\cite{Nagrani17} and out-of-domain on CommonBench~\cite{hintz24_spsc} and CN-Celeb~\cite{fan2019cnceleb}. Metrics are EER and minDCF with $P_{\text{tar}}{=}10^{-2}$, $C_{\text{miss}}{=}C_{\text{fa}}{=}1$. 

\subsection{Hardware and computational cost}

All experiments were conducted on the Jean Zay supercomputer. For consistency, all experiments were performed on the same GPU: an NVIDIA Tesla V100 32GB. Energy usage and runtime are recorded via the CEEMS infrastructure introduced in Section~\ref{sec:measurement_tools}, which collects node-level measurements and derives per-job power usage and CO\textsubscript{2}-eq emissions.

All experiments were conducted in France, where electricity had an average carbon intensity of approximately 50 gCO\textsubscript{2}/kWh, significantly lower than in countries heavily reliant on fossil fuels, such as the United States (345 gCO\textsubscript{2}/kWh) or China (536 gCO\textsubscript{2}/kWh).

\subsection{Performance of the models}
\label{sec:results}

Table~\ref{tab:Performance_EER} summarizes the results obtained with a family of ResNet architectures that vary along three structural dimensions: depth, channel width, and the distribution of residual blocks across stages. The depth ranges from 18 layers (9M parameters) to 419 layers (133M parameters). All models are evaluated on both in-domain and out-of-domain test sets.

For each system, the table reports the number of parameters, environmental-impact indicators (energy consumption and carbon footprint), and speaker verification performance in terms of EER and minDCF. Results are detailed for the in-domain datasets VoxCeleb1-O, VoxCeleb1-E, and VoxCeleb1-H, as well as for the out-of-domain datasets CommonBench and CN-Celeb. The average EER and average minDCF correspond to the mean values computed across all evaluation corpora.

\paragraphTwo{Depth vs. Accuracy}: increasing network depth (i.e., the number of hidden layers) generally improves performance, but with rapidly diminishing returns. From \emph{ResNet-18-D} to \emph{ResNet-419-D}, the average EER decreases from 4.11\% to 3.35\%, while the average minDCF drops from 0.273 to 0.218. However, most of the gains are observed between shallow models and mid-sized models: moving from \emph{ResNet-18-D} to \emph{ResNet-101-D/200-D} yields a significant improvement, whereas the transition from \emph{ResNet-200-D} to \emph{ResNet-419-D} results in only a marginal performance gain (from 3.44\% to 3.35\% in average EER). A similar trend is observed on out-of-domain datasets: all models exhibit degraded EER and minDCF on CommonBench and CN-Celeb, but the gap between \emph{ResNet-200-D} and \emph{ResNet-419-D} remains negligible, even though the latter requires more than four times the training energy cost. These results indicate that beyond a certain depth, adding more layers mainly increases energy consumption and carbon footprint without providing commensurate performance gains.

\paragraphTwo{Width vs. Accuracy}: beyond depth, we also investigate the impact of network width, measured by the number of channels in the convolutional layers. Increasing the number of channels tends to improve the average EER and average minDCF on in-domain datasets, but these gains do not consistently generalize to out-of-domain conditions. The \emph{ResNet-50-W2} model achieves better performance than \emph{ResNet-50-D} on VoxCeleb1-O/E/H and CN-Celeb, but performs worse on CommonBench. Similar trends were observed for deeper models such as \emph{ResNet-101-D}. In contrast, extreme width expansion is less effective: \emph{ResNet-50-W4} ($\sim$233M parameters) achieves an average EER of 3.77\%, whereas \emph{ResNet-50-D} achieves an average EER of 3.75\%.

Narrower variants, such as \emph{ResNet-50-W0.5} and \emph{ResNet-50-W0.25}, yield worse EER and minDCF than the \emph{ResNet-50-D} baseline. However, \emph{ResNet-50-W0.5} (average EER of 3.89\%) achieves performance comparable to that of \emph{ResNet-34-D} (average EER of 3.90\%), while using approximately three times fewer parameters and significantly less energy. This result highlights that a moderate reduction in network width can provide a favorable trade-off between performance and environmental impact.

\paragraphTwo{Stage distribution vs. Accuracy}: redistributing residual blocks across stages also affects performance. For \emph{ResNet-50-D} (average EER of 3.75\%), concentrating blocks in stages 2 and 3 improves both EER and minDCF: the \emph{ResNet-50-D$_{[2,8,2,2]}$} and \emph{ResNet-50-D$_{[2,6,6,2]}$} variants both reduce the average EER to 3.65\%, while the \emph{ResNet-50-D$_{[2,2,8,2]}$} variant achieves 3.68\%. In contrast, shifting residual blocks toward the first or the last stage degrades performance: \emph{ResNet-50-D$_{[8,2,2,2]}$} reaches an average EER of 3.80\%, and \emph{ResNet-50-D$_{[2,2,2,8]}$} reaches 3.91\%.

Overall, these results show that concentrating residual blocks in stages 2 and 3 of ResNet is more effective than allocating them to stages 1 or 4. Models following this strategy can reduce the number of parameters while matching or even surpassing the performance of the baseline model. This highlights the predominant role of stages 2 and 3 in the performance of speaker verification systems.

\subsection{Training-time, energy and carbon footprint}
\label{sec:energy}

Table~\ref{tab:Performance_EER} reports the energy consumption and carbon emissions for each ResNet variant.

\paragraphTwo{Energy consumption and carbon footprint}: 
Table~\ref{tab:Performance_EER} shows that energy use does not scale linearly with model size. For example, \emph{ResNet-34-D} (16M parameters) consumes 46.2 kWh, while \emph{ResNet-200-D} (61M) rises to 222.5 kWh. Beyond this point, the cost grows disproportionately: \emph{ResNet-419-D} (133M) reaches 896 kWh, $4\times$ \emph{ResNet-200-D}, for marginal EER and minDCF gains, emitting 13.7 kg CO\textsubscript{2}-eq even under France’s low-carbon grid (50 gCO\textsubscript{2}/kWh).

Focusing on \emph{ResNet-50-D} (17M, 52 kWh, 0.98 kg CO\textsubscript{2}), width scaling shows the extremes of this trade-off. Doubling width (\emph{ResNet-50-W2}, 61M) more than doubles energy demand (120.7 kWh, 1.66 kg CO\textsubscript{2}), while extreme widening (\emph{W4}, 233M) reaches 438 kWh and 6.98 kg CO\textsubscript{2}, rivaling \emph{ResNet-419-D}. Conversely, narrower variants (\emph{W0.5}, 5M; \emph{W0.25}, 2M) cut consumption to 40 and 34 kWh with emissions under 0.7 kg, but at the cost of much higher EER.

Stage-distribution models of \emph{ResNet-50-D} show a different profile: redistributing blocks can lower cost, although the parameter count may also change depending on how blocks are allocated across stages. For instance, \emph{ResNet-50-D$_{[2,2,8,2]}$} and \emph{ResNet-50-D$_{[2,2,2,8]}$} consume only 49–47 kWh, with emissions around 0.7 kg CO\textsubscript{2}, while mid-balanced variants (\emph{[2,8,2,2]} and \emph{[2,6,6,2]}) stay close to baseline at 52–53 kWh (0.77–0.80 kg). Thus, reallocating blocks to intermediate or later stages can slightly reduce energy consumption and carbon footprint, although the parameter count is not strictly constant across these variants. Overall, width scaling strongly amplifies environmental impact as parameters grow, whereas block concentration offers small but consistent efficiency gains at constant size.

\begin{figure}[htbp]
  \centering
  \includegraphics[width=\linewidth]{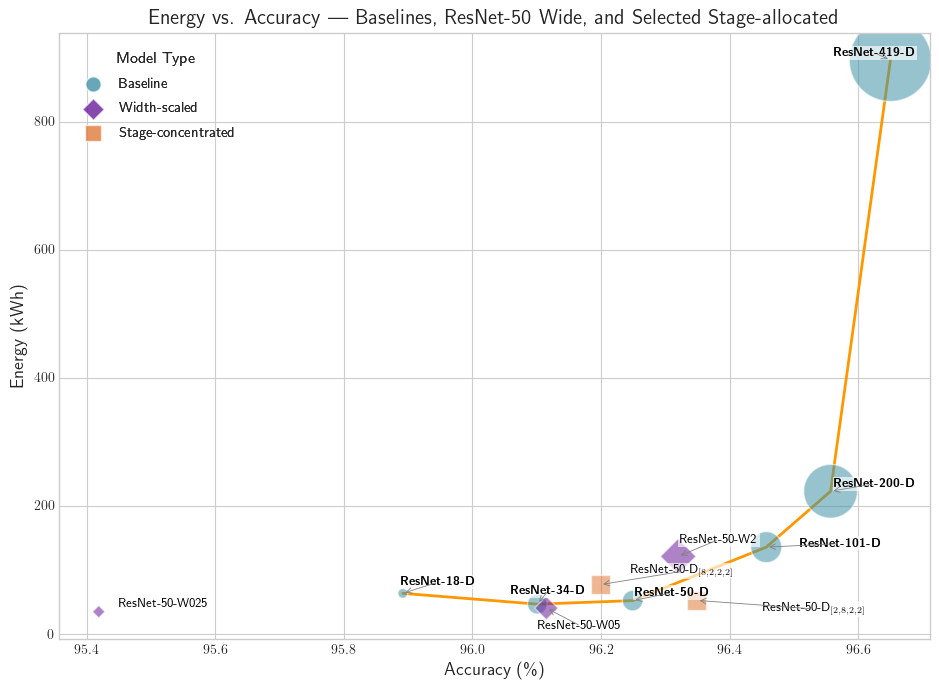}%
   \caption{Performance--energy trade-offs for ResNet-X-D architectures. The ``elbow'' is located around \emph{ResNet-101-D}/\emph{ResNet-200-D}, beyond which additional depth brings steep energy consumption for modest performance gains.}
  \label{fig:elbow}
\end{figure}

\paragraphTwo{Performance--energy trade-offs and sweet spot}: Figure~\ref{fig:elbow} plots a subset of representative models to highlight energy consumption vs. accuracy, revealing a clear ``knee’’ where extra capacity no longer yields proportional gains. \emph{ResNet-34-D} and \emph{ResNet-50-D} emerge as optimal sweet spots, delivering strong accuracy with energy consumption up to an order of magnitude lower than very deep models (\emph{ResNet-419-D}, 896 kWh).
Redistributing residual blocks toward the central stages improves efficiency: the stage-concentrated variant \emph{ResNet-50-D$_{[2,8,2,2]}$} achieves a better accuracy--energy balance than the baseline, although its parameter count is lower. 
Narrow models (e.g., \emph{ResNet-50-W0.25}, 2M) minimize energy consumption but suffer from degraded EER (4.58\%).

\paragraphTwo{Practical recommendations}: Based on the results presented above, several practical recommendations can be drawn for the design of energy-aware speaker verification systems. First, balanced mid-sized architectures such as \emph{ResNet-34-D} and \emph{ResNet-50-D} should be preferred for general-purpose deployments, as they provide the most favorable compromise between verification performance and environmental impact. Second, narrower variants such as \emph{ResNet-50-W0.5} are attractive for resource-constrained scenarios, since they substantially reduce energy consumption and carbon footprint while maintaining competitive accuracy. Third, very deep models beyond \emph{ResNet-101-D}/\emph{ResNet-200-D}, as well as extremely wide variants such as \emph{ResNet-50-W4}, should be avoided unless their marginal performance gains clearly justify their much higher environmental impact. Finally, reallocating residual blocks toward intermediate stages appears to be a more effective design strategy than concentrating capacity in the earliest or latest stages, as it can improve the overall performance--efficiency trade-off.

\subsection{Inference footprint}
\label{sec:inference}


Table~\ref{tab:inference_cost} shows the energy and emissions required to extract speaker embedding from the full CommonBench dataset (comprising over 9.7M pairs).

\begin{table}[H]
\centering

\resizebox{\columnwidth}{!}{
\begin{tabular}{@{}lcccccc@{}}
\toprule
\textbf{System} &
  \multicolumn{1}{c}{\textbf{EER}} &
  \multicolumn{1}{c}{\textbf{minDCF}} &
  \textbf{\begin{tabular}[c]{@{}c@{}}Energy \\ (kWh)\end{tabular}} &
  \textbf{\begin{tabular}[c]{@{}c@{}}CO$_2$ \\ (kg-eq)\end{tabular}} &
  \textbf{\begin{tabular}[c]{@{}c@{}}Duration\\ (HH:MM)\end{tabular}} &
  \textbf{Precision} \\ \midrule
ResNet-419 & -     & -     & -             & -               & -                 & -    \\
ResNet-200 & 3.456 & 0.329 & 8.51          & 0.1224          & 02:12:48          & FP32 \\
ResNet-101 & 3.502 & 0.325 & 5.03          & 0.0650          & 01:15:22          & FP32 \\
ResNet-50  & 3.814 & 0.345 & 3.11          & 0.0405          & 00:51:57          & FP32 \\
ResNet-34  & 3.935 & 0.359 & 2.67          & 0.0321          & 00:46:17          & FP32 \\
ResNet-18  & 4.181 & 0.372 & \textbf{1.94} & \textbf{0.0239} & \textbf{00:36:33} & FP32 \\ \midrule
ResNet-419 & 3.475 & 0.330 & 8.58          & 0.1116          & 03:26:48          & FP16 \\
ResNet-200 & 3.462 & 0.327 & 5.45          & 0.0685          & 01:53:11          & FP16 \\
ResNet-101 & 3.502 & 0.325 & 3.13          & 0.0344          & 01:09:28          & FP16 \\
ResNet-50  & 3.813 & 0.345 & 2.30          & 0.0253          & 00:53:37          & FP16 \\
ResNet-34  & 3.935 & 0.359 & 1.94          & 0.0234          & 00:44:18          & FP16 \\
ResNet-18  & 4.181 & 0.372 & \textbf{1.53} & \textbf{0.0184} & \textbf{00:37:48} & FP16 \\ \bottomrule
\end{tabular}

}
\caption{Inference-time performance, energy consumption, carbon emissions, and runtime of ResNet speaker verification models on CommonBench under FP32 and FP16 precision.}
\label{tab:inference_cost}
\end{table}

We report results in both standard full precision (FP32) and mixed precision (FP16) modes. We observe that EER and minDCF are practically the same for FP16 and FP32. Energy usage scales with model size, with \emph{ResNet-18} consuming as little as 1.94\,kWh in FP32 and just 1.53\,kWh in FP16.
We note that \emph{ResNet-419} was executed using only mixed precision (FP16), due to GPU memory limitations.

These results confirm that although inference costs are significantly lower than training, they remain non-negligible at scale. For large multilingual corpora like CommonBench, the difference between architectures is significant: \emph{ResNet-419} emits almost 6 times more CO\textsubscript{2} than \emph{ResNet-18} when using the same precision (FP16).

Switching to FP16 yields consistent gains across all models, with average energy savings of 25--35\% and reduced inference times while maintaining the same performance. This highlights mixed precision as an effective low-effort optimization for energy-aware deployment of speaker verification systems. For real-world deployments in resource-constrained or carbon-sensitive environments, such differences need careful consideration too when selecting a backbone architecture.

\section{Conclusions}
\label{sec:conclusions}

This paper analyzed the trade-offs between accuracy, energy consumption, and carbon footprint in ResNet-based speaker verification by varying depth, width, and stage distribution. Results show that although deeper models improve EER and minDCF, these gains diminish sharply beyond \emph{ResNet-101-D}/\emph{ResNet-200-D}, while energy consumption and carbon footprint increase disproportionately.

Width scaling showed that moderate expansion (W2) can provide small in-domain gains, whereas extreme widening (W4) leads to a disproportionately high environmental impact. Conversely, narrower variants such as \emph{ResNet-50-W0.5} and \emph{ResNet-50-W0.25} substantially reduce energy consumption, at the expense of absolute accuracy.

Stage-distribution models further showed that reallocating residual blocks toward intermediate stages (stages 2 and 3) can improve the performance--efficiency trade-off, whereas concentrating capacity in the earliest or latest stages degrades performance.

Overall, our results suggest the following practical recommendations: balanced mid-sized models such as \emph{ResNet-34-D} and \emph{ResNet-50-D} should be preferred for general-purpose SV systems; narrower variants are attractive for resource-constrained deployments; and very deep or extremely wide architectures should be avoided unless their marginal performance gains clearly justify their additional environmental impact.

\section{Acknowledgements}

This work was granted access to the HPC resources of IDRIS under the allocations AD011013257R4 and AD011016050R1 made by GENCI.

\bibliographystyle{IEEEtran}
\bibliography{Odyssey2026_BibEntries}

\end{document}